\documentclass[floatfix,twocolumn,prd,showpacs,preprintnumbers,amsmath,nofootinbib,amssymb,noeprint,superscriptaddress]{revtex4-1}
\usepackage{amsmath,amssymb,amsfonts,mathtools}
\usepackage{float}
\usepackage{graphicx}
\usepackage[colorlinks=true,citecolor=blue,urlcolor=blue,linkcolor=blue]{hyperref}
\usepackage{subfiles}
\usepackage{bbold}

\newcommand{\tr}{{\rm tr} }

\begin{document}

\title{Universality of the Collins-Soper kernel in lattice calculations}

\author{Hai-Tao Shu}
\affiliation{Institut f\"ur Theoretische Physik, Universit\"at Regensburg, D-93040 Regensburg, Germany}

\author{Maximilian Schlemmer}
\affiliation{Institut f\"ur Theoretische Physik, Universit\"at Regensburg, D-93040 Regensburg, Germany}

\author{Tobias Sizmann}
\affiliation{Institut f\"ur Theoretische Physik, Universit\"at Regensburg, D-93040 Regensburg, Germany}

\author{Alexey Vladimirov}
\affiliation{Departamento de F\'{i}sica Te\'{o}rica \& IPARCOS, Universidad Complutense de Madrid, E-28040 Madrid,
Spain}

\author{Lisa Walter}
\affiliation{Institut f\"ur Theoretische Physik, Universit\"at Regensburg, D-93040 Regensburg, Germany}

\author{Michael Engelhardt}
\affiliation{Department of Physics, New Mexico State University, Las Cruces, New Mexico 88003, USA}

\author{Andreas Sch\"afer}
\affiliation{Institut f\"ur Theoretische Physik, Universit\"at Regensburg, D-93040 Regensburg, Germany}

\author{Yi-Bo Yang}
\affiliation{CAS Key Laboratory of Theoretical Physics, Institute of Theoretical Physics, Chinese Academy of Sciences, Beijing 100190, China}
\affiliation{School of Fundamental Physics and Mathematical Sciences, Hangzhou Institute for Advanced Study, UCAS, Hangzhou 310024, China}
\affiliation{International Centre for Theoretical Physics Asia-Pacific, Beijing/Hangzhou, China}
\affiliation{School of Physical Sciences, University of Chinese Academy of Sciences,
Beijing 100049, China}

\begin{abstract}
The Collins-Soper (CS) kernel is a nonperturbative function that characterizes the rapidity evolution of transverse-momentum-dependent parton distribution functions (TMDPDFs) and wave functions. In this paper, we calculate the CS kernel for pion and proton targets and for quasi-TMDPDFs of leading and next-to-leading power. The calculations are carried out on the CLS ensemble H101 with dynamical $N_f=2+1$ clover-improved Wilson fermions. Our analyses demonstrate the consistency of different lattice extractions of the CS kernel for mesons and baryons, as well as for twist-two and twist-three operators, even though lattice artifacts could be significant. This consistency corroborates the universality of the lattice-determined CS kernel and suggests that a high-precision determination of it is in reach.


\end{abstract}

\preprint{IPARCOS-UCM-23-027}
\maketitle

\section{Introduction} 
The description of the internal structure of hadrons is a fundamental problem of QCD. The present description of high-energy processes is founded on factorization theorems, which express the cross sections of reactions in terms of calculable perturbative parts and universal nonperturbative functions.
In the modern era, the emphasis of studies is shifting towards multidimensional observables \cite{Amoroso:2022eow, AbdulKhalek:2022hcn} and multidimensional parton distributions, such as transverse momentum dependent parton distribution functions (TMDPDFs) \cite{Angeles-Martinez:2015sea}.
TMDPDFs encode information about the 3D parton momenta inside a hadron. TMDPDFs are assumed to be universal in the perturbative domain (small $b$ in Eq.(\ref{eq_evo})), i.e., to depend on the types of parton and hadron but not on the process. This universality is the cornerstone of the factorization approach. In the case of TMDPDFs, it has been only indirectly confirmed by many phenomenological extractions, which utilize multiple processes \cite{Scimemi:2017etj,  Bertone:2019nxa, Vladimirov:2019bfa, Scimemi:2019cmh, Bacchetta:2019sam, Bacchetta:2022awv, Bury:2020vhj, Echevarria:2020hpy}. 

The evolution of TMDPDFs with the rapidity scale $\zeta$ is described by 
\begin{equation}
\label{eq_evo}
2\zeta \frac{\rm{d}}{\rm{d} \zeta}\ln \Phi_{f/h}(x, b, \mu, \zeta) = K(b, \mu),
\end{equation}
which also provides the simplest way to access the CS kernel $K(b, \mu)$--the topic of this study.
For large $b$, where there exists only little experimental data, the status of the universality of the CS kernel is questionable. It can at present only be tested by lattice simulations like ours. $\Phi_{f/h}$ is a TMDPDF of flavor $f$ in hadron $h$ with $x$ being the longitudinal momentum fraction, and $b$ being the transverse distance, at scale $\mu$. The evolution equation Eq.~(\ref{eq_evo}) is predicted to hold for TMDPDFs of any kind, including also twist-3 TMDPDFs, as was derived recently in \cite{Vladimirov:2021hdn, Ebert:2021jhy, Rodini:2022wki} and is verified for the first time by lattice calculation in this work. In fact, the CS kernel is one of the most fundamental nonperturbative functions in QCD since it describes the interaction of a parton with the QCD vacuum \cite{Vladimirov:2020umg} and appears in the description of many types of processes, including inclusive ones \cite{Collins:2011zzd}, exclusive ones \cite{Echevarria:2022ztg}, and jet-production \cite{Neill:2016vbi}. Being a vacuum-determined function, the CS kernel obeys a stronger universality -- it is independent of any quantum numbers except the color representation of the probe (quark or gluon). The confirmation of this universality for the CS-kernel is of fundamental importance for QCD. 

Traditionally, the CS kernel is determined from fits of scattering data, along with TMDPDFs; see \cite{Landry:1999an, Scimemi:2019cmh, Bacchetta:2019sam} for examples. However, this approach requires assumption of a functional form and, thus, is biased. Recently a number of more direct ways were proposed. All these methods suggest to determine the CS kernel from the ratio of properly constructed observables -- cross sections \cite{BermudezMartinez:2022ctj} or quasi-TMDPDFs \cite{Ebert:2018gzl, Ji:2019sxk, Vladimirov:2020ofp}. The latter can be achieved by lattice QCD simulations, which have been done in Refs. \cite{Ebert:2018gzl, LatticeParton:2020uhz, Schlemmer:2021aij, Li:2021wvl, LPC:2022ibr}.
So far, all simulations were done for unpolarized quasi-TMDPDFs of the proton. The only exception is \cite{Schlemmer:2021aij}, where also polarized quasi-TMDPDFs are used. 

In this paper, we present a new set of lattice computations of the CS kernel. We compare results for four basically independent calculations, namely for proton and pion targets, and twist-two and twist-three quasi-TMDPDF operators. The agreement between the results confirms universality of the CS kernel and further constrains its form. This is a proof of principle. The precision of such tests will improve continuously in the future.  

\begin{figure}[tb]
\includegraphics[width=.4\textwidth]{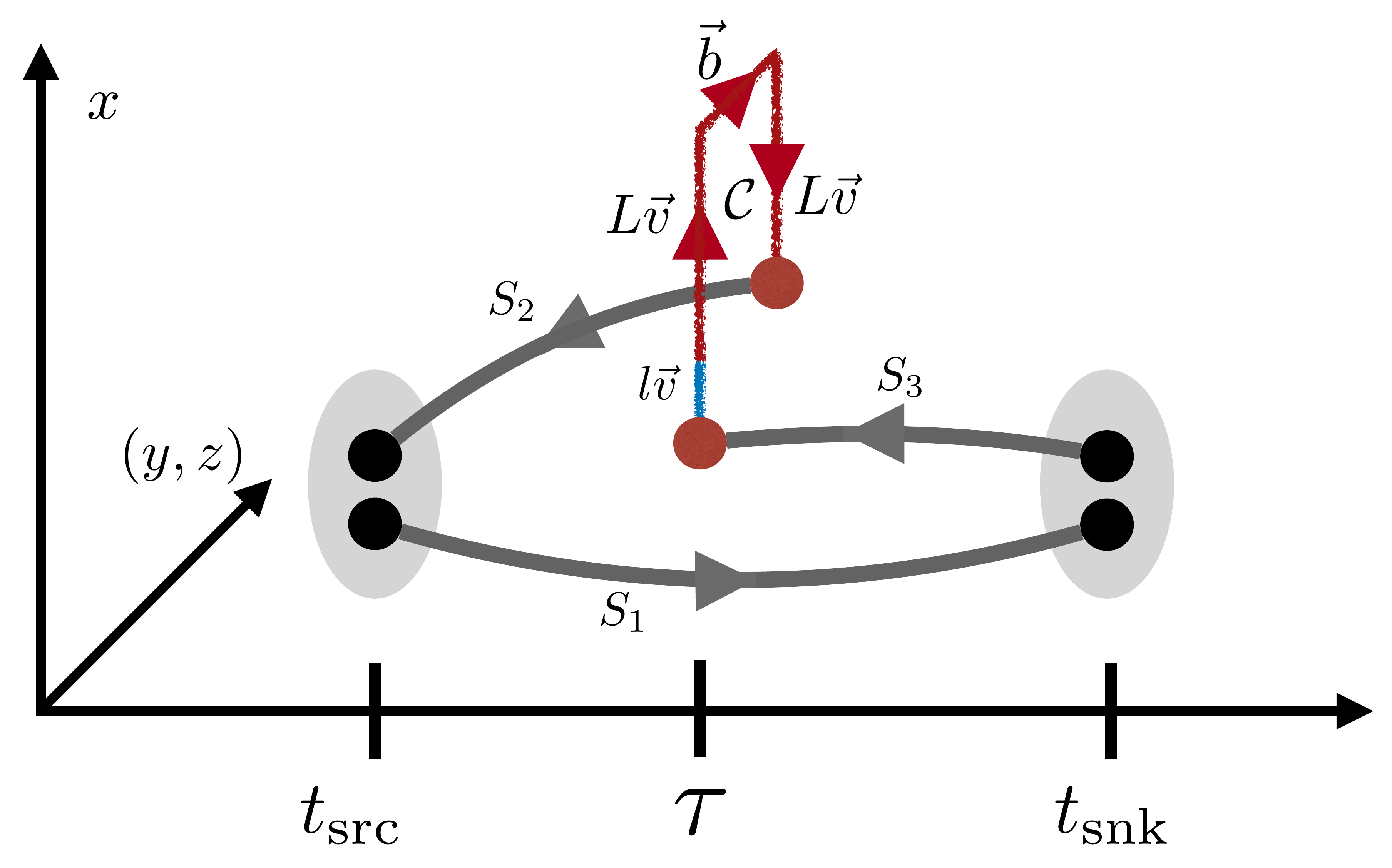}
\caption{Illustration of the pion matrix element, see Eq.(\ref{th:W-def}). The two external pion states are shown as gray ovals. The nonlocal quark current is shown in red, made up of a quark-antiquark pair (the red points) connected by a staple-shaped gauge link $\mathcal{C}$.
The matrix element is calculated on the lattice using the sequential source method. The sequential source is constructed using propagators $S_1$ and $S_3$. For the proton an additional direct propagator from source to sink is needed.
} 
\label{sketch}
\end{figure} 

\section{Theoretical framework}
We consider the following matrix element (quasi-TMDPDF)
\begin{equation}
\label{th:W-def}
\begin{split}
&W^{[\Gamma]}_{f/h}(b;\ell, L;v,P, S;\mu)=\\
&\langle h(P,S)|\bar{q}_f(b+\ell v)\Gamma\;\mathcal{U}[\mathcal{C}(\ell, v, b, L)]\;q_f(0)|h(P,S)\rangle,
\end{split}
\end{equation}
where $|h(P,S)\rangle$ is a single-hadron state with momentum $P$ ($P^{\mu}=(E_P, P_x, 0, 0)$ in this study) and spin $S$ (which is suppressed in the following for brevity). $\Gamma$ is a Dirac matrix and $f$ indicates the flavor of the quark field. The staple-shaped Wilson link $\mathcal{U}$ is of length $L$ stretching in the direction $v^\mu=(0,\pm 1,0,0)$ and of width $\vec{b}$ pointing in a transverse spatial direction. 
The quark-antiquark pair connected by the Wilson link is positioned in the same imaginary time slice. The offset of the quark-antiquark pair along $v$ is denoted by $\ell$. The structure of the matrix element is sketched in Fig.~\ref{sketch}.

At large $L$ and hadron momentum, it is assumed that the matrix element (\ref{th:W-def}) can be factorized \cite{Ebert:2018gzl, Vladimirov:2020ofp, Ji:2019sxk, Ji:2021znw, Ebert:2022fmh} after having been transformed to momentum space. The structure of the factorization theorem crucially depends on $\Gamma$. In particular, for $\Gamma=\{\gamma^0, \not \!v, ...\}$ (the complete set can be found in Ref. \cite{Rodini:2022wic}) one has the so-called leading power (LP) expression 
\begin{equation}
\label{th:W-fac}
\begin{split}
&W^{[\Gamma]}_{f/h}(x;b;P;\mu)=\Big{(}\frac{2|x|(P^+)^2}{\zeta}\Big{)}^{K(b,\mu)/2}\\
&\ \ \ \ \ \ \ \ \ \ \times\mathbb{C}_H(xP^+, \mu)\Phi^{[\Gamma]}_{f/h}(x,b;\mu,\zeta)+\mathcal{O}(\lambda^2),
\end{split}
\end{equation}
where $P^+=(E_P+P_x)/\sqrt{2}$ and $\Phi$ is the physical TMDPDF and $\mathbb{C}_H$ is the coefficient function. The coefficient functions $\mathbb{C}_H$ are known at next-to-leading order (NLO) in the QCD coupling constant \cite{Ebert:2018gzl, Vladimirov:2020ofp, Ji:2019sxk}. The variable $x$ is the momentum fraction, Fourier-conjugate to $\ell P_x$, and $M$ is the mass of the hadron.  $\mathcal{O}(\lambda^2)$ contains various power-suppressed terms,
\begin{eqnarray}
\label{hierarchy}
\mathcal{O}(\lambda^2)=\mathcal{O}\Big(\frac{M^2}{(xP^+)^2},\frac{1}{(bP^+)^2},\frac{b}{L},\frac{1}{ML}\Big).
\end{eqnarray}

The left-hand side of Eq.~(\ref{th:W-fac}) is independent of $\zeta$. Thus, the ratio of quasi-TMDPDFs which differ only in their momenta is simply
\begin{eqnarray}
\nonumber
&&R^{[\Gamma]}(x,b,\mu;P_1,P_2)=\frac{W^{[\Gamma]}_{f/h}(x,b;P_1,S;\mu)}{W^{[\Gamma]}_{f/h}(x,b; P_2,S;\mu)}
\\\label{th:ratio-cs-kernel}
&&\quad =
\left(\frac{P_1^+}{P_2^+}\right)^{K(b,\mu)}\frac{\mathbb{C}_H(xP^+_1,\mu)}{\mathbb{C}_H(xP^+_2,\mu)}+\mathcal{O}(\lambda^2).
\end{eqnarray}
Inverting this relation one determines the CS-kernel. 

One should note that for this procedure, a Fourier transformation of the quasi-TMDPDF from coordinate space ($\ell P_x$) to momentum-fraction space ($x$) is required. Such a transformation requires model assumptions concerning the tail of the quasi-TMDPDF \cite{LPC:2022ibr}, which introduces additional systematic uncertainty. Furthermore, in \cite{LatticePartonLPC:2023pdv} it is shown that the CS kernel extracted in this way can be sensitive to $x$-dependent higher-twist effects, which are much stronger than those in the TMD wave function case. This complication is avoided when the ratio of the first Mellin moments ($\ell=0$, accordingly $x$ is suppressed in the following) of the two quasi-TMDPDFs is considered \cite{Vladimirov:2020ofp}. It reads
\begin{equation}
\label{th:ratio-cs-kernel-simple}
R^{[\Gamma]}(P_1,P_2;b)=\left(\frac{P_1^+}{P_2^+}\right)^{K(b,\mu)}\mathbf{r}^{[\Gamma]}(b,\mu;P_1,P_2),
\end{equation}
where $\mathbf{r}$ is \cite{Vladimirov:2020ofp}
 \begin{equation}
 \label{th:r-pert}
 \begin{split}
&\mathbf{r}^{[\Gamma]}(b,\mu;P_1,P_2)
=1+4C_F\frac{\alpha_s(\mu)}{4\pi}\ln\left(\frac{P_1^+}{P_2^+}\right)\\
&\times \Big[1-\ln\left(\frac{2P_1^+P_2^+}{\mu^2}\right)-2 \mathbf{M}^{[\Gamma]}(b,\mu)\Big].
 \end{split}
\end{equation}
The function $\mathbf{M}$ contains the residual terms of the perturbative expansion, and depends on the quantum numbers of the quasi-TMDPDF. A key argument underlying this method is that the function $\mathbf{M}^{[\Gamma]}(b,\mu)$ is almost independent of $b$. This assumption is based on the weak correlation of $b$ and $x$ dependencies of TMDPDFs, which has been verified by fitting experimental data with the unpolarized TMDPDFs of proton and pion \cite{Scimemi:2019cmh, Vladimirov:2019bfa}. The value of $\mathbf{M}$ can be found by comparing Eq.~(\ref{th:r-pert}) and its value in perturbation theory at $b\sim 1$ GeV$^{-1}$ and $\mu_0=2$ GeV (where both perturbation theory and the factorization theorem should be valid). For details see Ref. \cite{Schlemmer:2021aij}. This method is much simpler than evaluating Eq.~(\ref{th:ratio-cs-kernel}) but cannot be improved beyond NLO.

The description of the cases $\Gamma=\{\mathbb{1},\gamma^5, ...\}$ requires the next-to-leading power (NLP) factorization theorem \cite{Rodini:2022wic}. NLP factorization has a much more involved form and expresses a single quasi-TMDPDF by a sum of various physical TMDPDFs and new lattice-related nonperturbative functions $\Psi_{21}(b)$ and $\Psi_{12}(b)$ \cite{Rodini:2022wic}.
For particular combinations of $\Gamma$ and polarization, the NLP factorization simplifies to the form of Eq.~(\ref{th:W-fac}) (with a different coefficient function). In these cases, one can use Eq.~(\ref{th:ratio-cs-kernel-simple}) to determine the CS-kernel (note that Eq.~(\ref{th:ratio-cs-kernel}) is not helpful due to the $x$-dependence of $\mathbb{C}_H$ at NLP). These simple cases include $\Gamma=\mathbb{1}$ for the TMDPDF $e(x,b)$.

\section{Lattice calculation}
\label{Lattice_calculation}
The matrix element Eq.~(\ref{th:W-def}) can be calculated as the ratio of a three-point and a two-point function on the lattice, 
\begin{equation}
\label{def_W_lattice}
   W^{[\Gamma]}=2E_P\lim_{0\ll \tau \ll t}\frac{C^{\Gamma}_{\mathrm{3pt}}(\vec{P},\mathcal{C},t,\tau,\Gamma)}{C_{\mathrm{2pt}}(\vec{P},t)},
\end{equation}
where $E_P$ is the energy of the hadron extracted from the two-point function. In the continuum limit, the lattice definition Eq.~(\ref{def_W_lattice}) reproduces the continuum definition, see e.g. Ref.~\cite{Musch:2010ka}. The parameters $t$ and $\tau$ are the source-sink separation and the temporal distance between the source and the inserted nonlocal quark current. The three-point function is defined as 
\begin{equation}
\label{eq:3pt-def-lat}
C^{\Gamma}_{3\mathrm{pt}}(\vec{P},\mathcal{C},t,\tau,\Gamma) \equiv \left\langle \tr \left\{\Gamma_S \mathcal{O}(\vec{P},t)\ J^\Gamma(\mathcal{C},\tau)\ \overline{\mathcal{O}}(\vec{P},0) \right\} \right\rangle
\end{equation}
and calculated using the sequential source method \cite{Martinelli:1988rr} with hadron interpolator $\mathcal{O}(\vec{P},t)$. The nonlocal quark current reads
\begin{equation}
\label{eq:def-quark-bilinear}
J^\Gamma(\mathcal{C},\tau) \equiv \bar{q}(b,\tau) \Gamma \mathcal{U}[\mathcal{C}(v, b, L)] q(0,\tau),
\end{equation}
where $q$ can be either up or down quark, and $\mathcal{U}[\mathcal{C}(v, b, L)]$ is the staple-shaped Wilson link in Fig.\ref{sketch}. The two-point function is 
\begin{equation}
\label{eq:2pt-def}
C_{2\mathrm{pt}}(\vec{P},t) \equiv \left\langle \tr \left\{ \Gamma_S \mathcal{O}(\vec{P},t)\overline{\mathcal{O}}(\vec{P},0) \right\} \right\rangle,
\end{equation}
where $\Gamma_S=(1+\gamma_4)(1-i\gamma_2\gamma_1)/2$ is needed for the proton to project out the desired parity and spin. For the pion $\Gamma_S$ is not necessary and is set to unity. We adopt HYP smearing for the gauge links \cite{Hasenfratz:2001hp}
and use momentum smearing \cite{Bali:2016lva} to improve the signal. We analyse the CLS ensemble H101 generated using $N_f=2+1$ flavors of clover-improved Wilson fermions \cite{Bruno:2014jqa}.  The lattice setup is summarized in Table \ref{Tab:setup}. 

\begin{table}[t]
\centering
\caption{Lattice setup used in this study.}
\label{Tab:setup}
\begin{tabular}{cclccccc}
\hline
\hline
Ensemble ~~& $a$[fm] & \ \!$N_{\sigma}^3\times\  N_{\tau}$   & $m^{sea}_{\pi}$  & \#Configuration \\
\hline
H101  ~~& 0.0854  ~~& $32^3\times$~ \!96  ~~& 422 MeV         ~~ & 2016  ~   \\
\hline
\hline
\end{tabular}
\end{table}

The analysis for the proton reuses the data generated in \cite{Schlemmer:2021aij}, where the source-sink separation is $t_{\mathrm{snk}}-t_{\mathrm{src}}=11a$ and the valence quark is the same as the sea quark. In the pion case, the simulation with the same setup is much noisier. To reduce the noise, we use a heavier valence quark corresponding to $m^{val}_{\pi}$=686 MeV. We do not expect a substantial mass dependence of the resulting CS kernel. In fact, in the physical limit, at large boost factors, the CS kernel depends only weakly on the quark mass, see, e.g., Ref. \cite{Li:2021wvl}. Besides, we use a smaller source-sink separation (9$a$) to further increase the signal for the three- and two-point functions.
   
For the pion, we fit the ratio to a constant in the interval $\tau \in [4a, 6a]$, where the excited states are suppressed. The simulation has been done for six momenta $P_1\in \{0,1,2,3,4,5\}\frac{2\pi}{aN_{\sigma}}$, but only the first three nonzero momenta have good enough signal/noise ratio to be processed further. We have confirmed that the extracted energies respect the dispersion relation $E_P=\sqrt{M^2+P_1^{2} }$ within statistical errors. We have also confirmed that our results respect the charge conservation condition $C^{[\gamma^0]}_{\mathrm{3pt}}(t,\tau,\gamma^0)/C_{\mathrm{2pt}}(t)=1/Z_V$, where $Z_V$ is the renormalization constant for the quark current in the vector channel~\cite{Bali:2020lwx} and $C^{[\gamma^0]}_{\mathrm{3pt}}(t,\tau,\gamma^0)$ is the local three-point function. For nonlocal correlations, we consider transverse separations in the $y$- or $z$-direction (or a combination of both), with lengths $\{1, \sqrt{2}, 2,..., 8, 6\sqrt{2}, 9\}a$. The size of the staple-link $L$ is taken as large as possible under the conditions that Eq.~(\ref{hierarchy}) is small and the signal for the three-point function is acceptable.

\begin{figure*}[t]
\centerline{
\includegraphics[width=.5\textwidth]{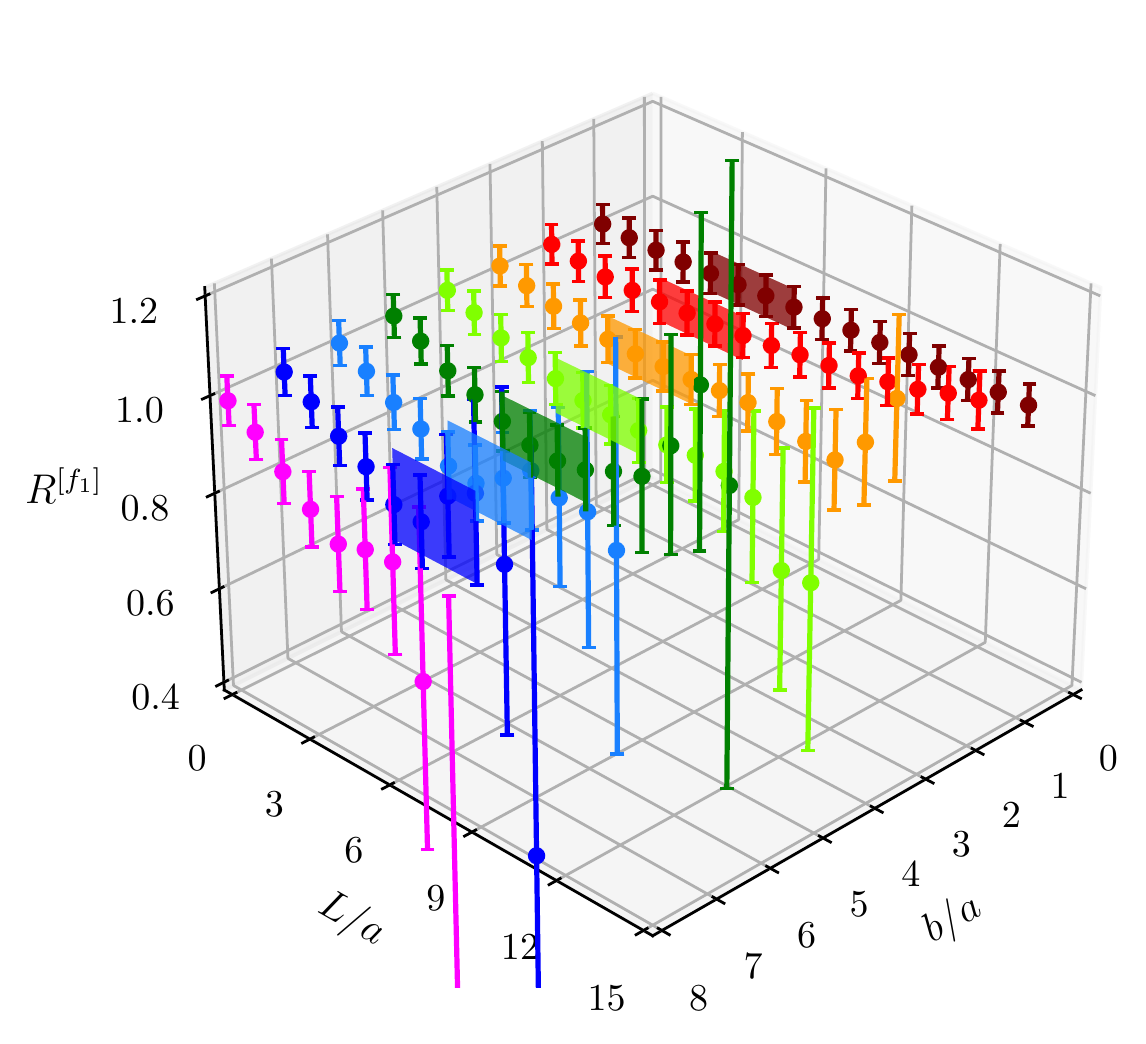}
\includegraphics[width=.5\textwidth]{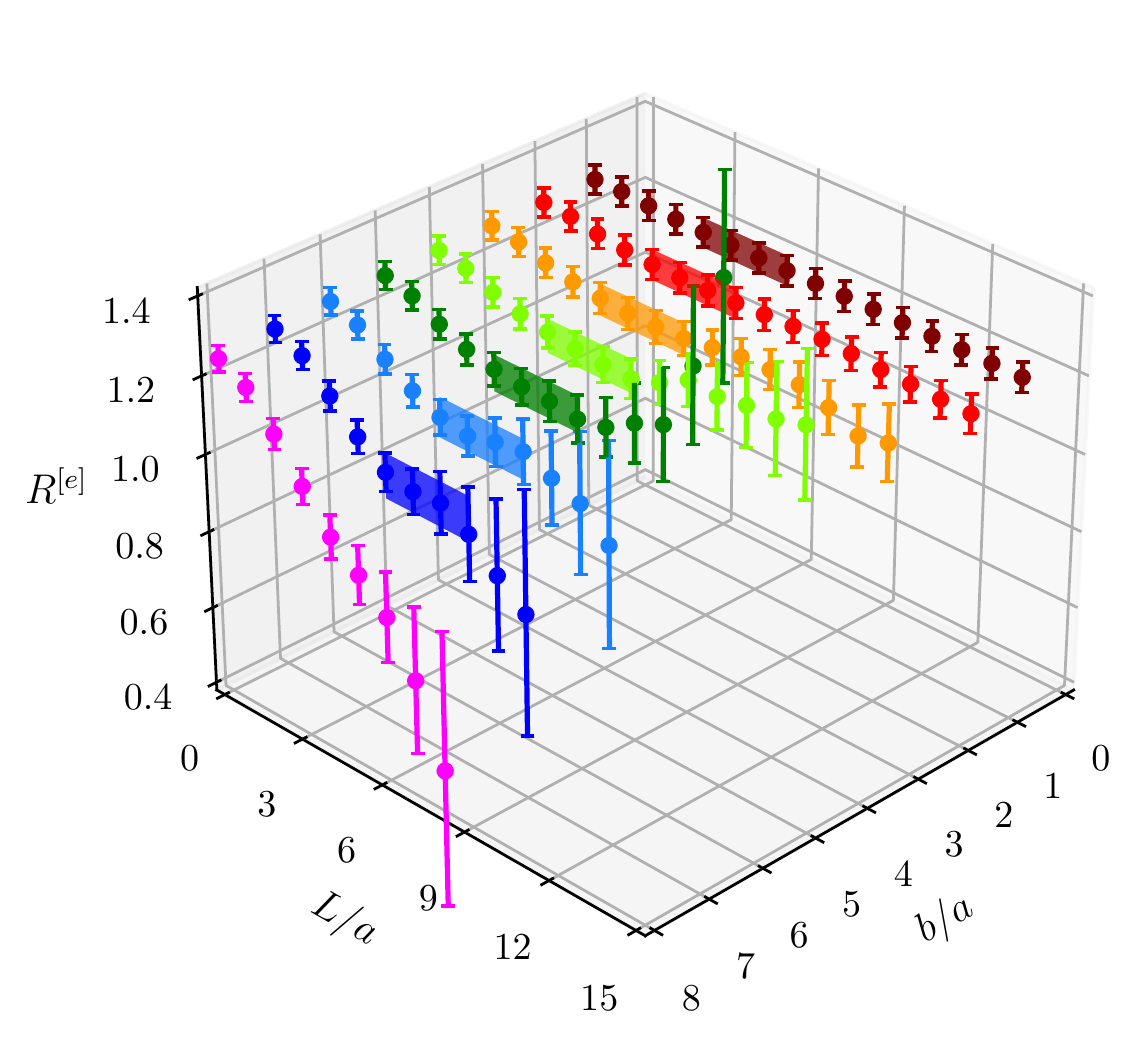}
}
\caption{Lattice results for the ratios $R^{[f_1]}$ and $R^{[e]}$ in the pion case from the momentum pair $\{P_1^+,\  P_2^+\}=\{2.07,\ 1.48\}\ \mathrm{GeV}$ at different transverse separations $b$. The colored bands indicate the results of constant fits in the $L$-interval $[4a, 7a]$.} 
\label{fig:ratio}
\end{figure*} 

As detailed in Refs.~\cite{Musch:2010ka,Musch:2011er,Engelhardt:2015xja,Yoon:2017qzo}, the hadron matrix elements Eq.~(\ref{def_W_lattice}) in different channels (for different $\Gamma$) can be parameterized using invariant amplitudes $\tilde{A}_i$ and $\tilde{B}_i$. For the pion this parametrization is \cite{Engelhardt:2015xja}
\begin{equation}
\begin{split}
&W^{[\gamma^0]}=E_P\tilde{A}_2,\\
&W^{[\gamma^1]}=P_x\tilde{A}_2+M^2\frac{v^{[1]}}{(v\cdot P)}\tilde{B}_1,\\
&W^{[\mathbb{1}]}=M\tilde{A}_1.
\end{split}
\label{explicit}
\end{equation}
Thus the relevant amplitudes can be obtained from a combination of matrix elements in the three channels $\gamma_0$, $\gamma_1$ and $\mathbb{1}$. Combining the available amplitudes which satisfy the LP ($f_1$) and NLP ($e$) factorization theorem \cite{Rodini:2022wic}, we obtain 
\begin{equation}
\begin{split}
&f_1(b^2, P^+)=P^+ \big{(}\tilde A_2(b^2)+M^2 \frac{v^+}{(v\cdot P) P^+}\tilde B_1(b^2)\big{)},\\
&e(b^2, P^+)=P^+ \tilde A_1(b^2).
\end{split}
\end{equation}

\section{Numerical results}
\label{Numerical_result} 
\begin{figure*}[t]
\centerline{
\includegraphics[width=.5\textwidth]{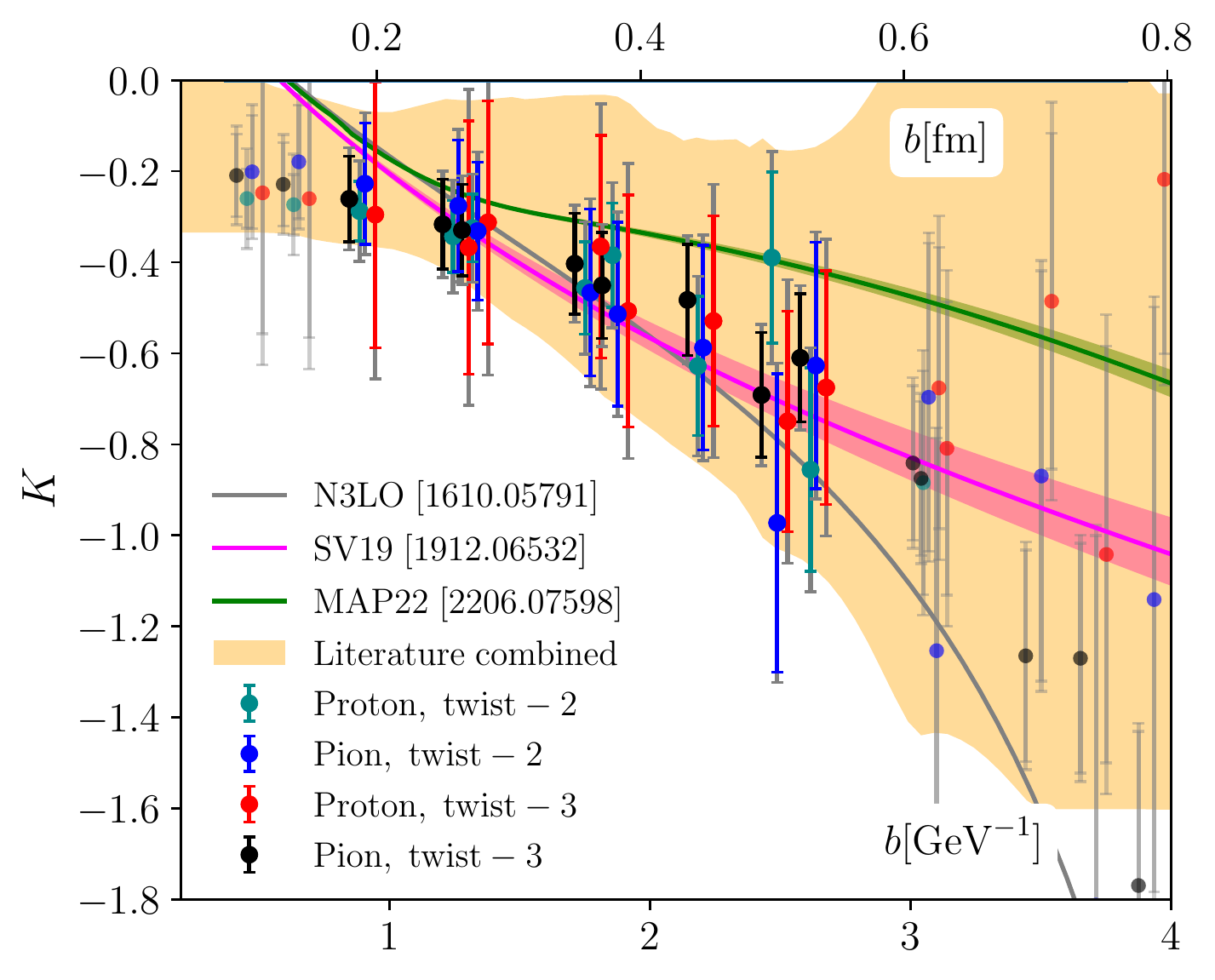}
\includegraphics[width=.5\textwidth]{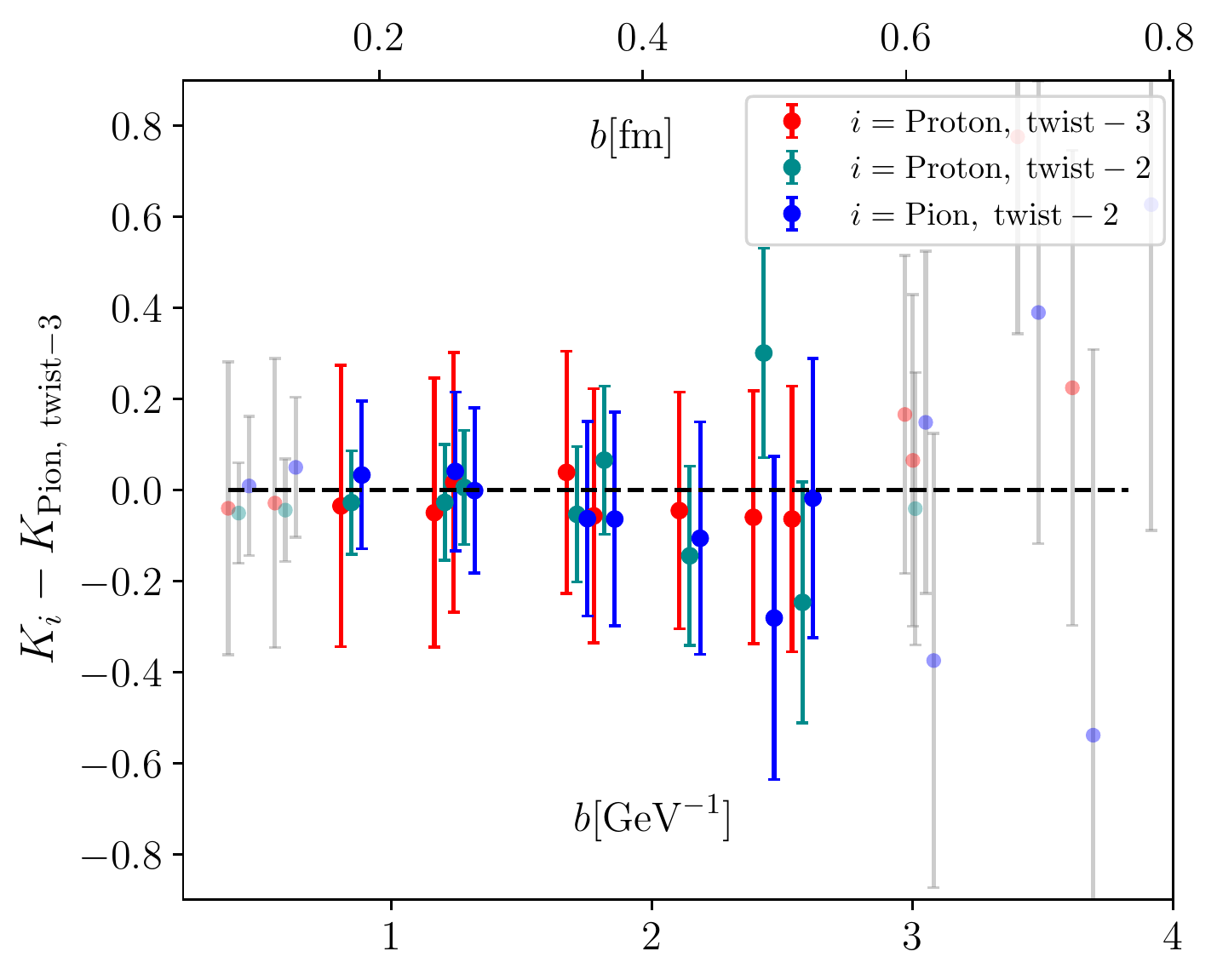}
}
\caption{Left: comparison of the CS kernel obtained in this work to the 3-loop perturbative calculation \cite{Li:2016ctv, Vladimirov:2016dll},  two phenomenological extractions, SV19 \cite{Scimemi:2019cmh} and MAP22 \cite{Bacchetta:2022awv} and the ``Literature combined" result that summarizes previous lattice extractions \cite{Li:2021wvl, LatticeParton:2020uhz, LPC:2022ibr, Shanahan:2020zxr, Shanahan:2021tst}. The outer error bars denote the possible lattice artifacts estimated in a way described in Appendix \ref{app:latt-artifact}. They are not shown in other figures. Right:  the differences between the most accurate result, which is the twist-3 pion case and the other  three extractions. For details see the main text. The points are slightly shifted horizontally for better visibility in both panels.} 
\label{fig:cskernel_final}
\end{figure*}

The momentum pair $P_x=\{2,3\}\times \frac{2\pi}{aN_{\sigma}}$, used for both pion and proton, corresponds to $P^-/P^+=\{0.13,\ 0.07\}$ for the pion and $P^-/P^+=\{0.25,\ 0.14\}$ for the proton, such that $M^2/(xP^+)^2$ is not large in Eq.~(\ref{hierarchy}). We also analyzed the momentum pair $P_x=\{1,2\}\times \frac{2\pi}{aN_{\sigma}}$ for the pion. This case has larger systematic uncertainties due to the power corrections in Eq.~(\ref{hierarchy}) which we cannot reliably quantify. On the other hand,
it has much smaller statistical uncertainties and we do not know for which combination the total uncertainty is larger. We present, therefore, our final results in Fig.~\ref{fig:cskernel_final} for the momenta $P_1/P_2=3/2$, i.e., with the larger statistical errors, and give the analog for $P_1/P_2 =2/1$ in Appendix \ref{app:momentum-eff}. $1/(bP^+)^2\ll 1$ for Eq.~(\ref{hierarchy}) is fulfilled if $b\gg \{1.8a, 1.1a\}$, implying that our extracted CS kernel is valid at $b>0.15$ fm. Finally, to make $b/L$ and $1/(ML)$ small, $L$ is chosen as large as possible. We observed  plateaus in the interval $[4a, 7a]$ for both $f_1$ and $e$ at all values of $b$ for the pion, see Fig.~\ref{fig:ratio}. The analogous figures for the proton are presented in Appendix \ref{app:more-R}. Still larger values of $L$ can also be included into the fit, but have negligible impact on the fit's quality. To increase the statistics and reduce systematic uncertainties, we have combined the data with $L$ pointing into the positive and negative $v$ directions.

The value of the constant $\mathbf{M}$ is determined following the procedure described in Ref.~\cite{Schlemmer:2021aij}. We use the reference transverse separation $b_0=3a=$\,0.26 fm\,=\,1.3 GeV$^{-1}$, for which the value of the CS kernel is safely known from perturbative computations \cite{Echevarria:2012pw, Li:2016ctv, Vladimirov:2016dll, Moult:2022xzt} and from phenomenological extractions \cite{Scimemi:2019cmh, Bacchetta:2022awv} (all agree with each other up to small corrections). At the same time, the terms in Eq.~(\ref{hierarchy}) are small. We normalize the value of the CS kernel at this point (explicitly, we use the values of the phenomenological extraction SV19 \cite{Scimemi:2019cmh} at N$^3$LO). Our estimate for $\mathbf{M}$ is -0.83(0.73) for $f_1$(pion), -4.98(0.61) for $e$(pion), -0.57(0.34) for $f_1$(proton), and -1.04(1.32) for $e$(proton).A cross check on the determination of $\mathbf{M}$ using other lattice CS kernel data can be found in Appendix \ref{app:check-M}. The uncertainty in the estimation of $\mathbf{M}$ results in a fully correlated uncertainty for the CS kernel $\delta K$. The values of $\delta K$ are $\{0.15,\ 0.10,\ 0.08,\  0.27\}$ for $f_1$(pion), $e$(pion), $f_1$(proton) and $e$(proton), correspondingly. Note that $\delta K$ is not shown in Figs.~\ref{fig:cskernel_final} and the figures in the 
appendices as it is dominated by the statistical uncertainty at the normalization point and, thus, cannot be simply added as an independent uncertainty.

The resulting values of the CS kernel are plotted in Fig.~\ref{fig:cskernel_final}. In the left plot, we compare our high momentum results for the CS kernel with two phenomenological extractions, SV19 \cite{Scimemi:2019cmh} and MAP22 \cite{Bacchetta:2022awv}, one 3-loop perturbative calculation \cite{Li:2016ctv, Vladimirov:2016dll} and a ``Literature combined" result shown as a yellow band that summarizes previous lattice calculations \cite{Li:2021wvl, LatticeParton:2020uhz, LPC:2022ibr, Shanahan:2020zxr, Shanahan:2021tst} in a way described in Appendix \ref{app:error-combine}. In the right plot we show the difference of the other three extractions performed in this study to the most accurate one, obtained at twist-3 in pion states. 

All lattice results display qualitatively similar behavior. The differences between them are probably mainly due to systematic effects, since most calculations differ in important aspects. For example, the computation \cite{LPC:2022ibr} is based on 1-loop matching while the computation in \cite{LatticeParton:2020uhz} is based on tree-level matching. This observation underlines the relevance of our results: Various systematic effects should differ markedly between pion and proton as well as between twist-2 and twist-3. Therefore, the close agreement of our four sets of data for the CS kernel does not only confirm its universality and the results of  \cite{Vladimirov:2021hdn, Ebert:2021jhy, Rodini:2022wki}, but it suggests also that the uncertainties are still dominated by statistics.

Note also that, since pion states possess higher symmetry, fewer amplitudes are involved in the parametrization for the pion than for the proton, leading to reduced uncertainties when solving for the amplitudes. Therefore, calculating the CS-kernel for a pion as we pioneered with this paper should be especially reliable.

In Fig.~\ref{fig:cskernel_final},  the data points with $b>7a$ are plotted with light colors and gray error bars to indicate that the extracted values suffer from uncontrolled systematic uncertainties. In this region, plateaus of $R$ are not reached even for the largest values of $L$, as can be seen in Fig.~\ref{fig:ratio}. Additionally, the points at small $b<0.8$ GeV$^{-1}$ are contaminated by power corrections $\sim b^{-2}$. Therefore, our main results are the points in the intermediate region.

\section{Conclusion}
\label{Conclusion}
We have extracted the CS kernel from the first Mellin moment of twist-2 and twist-3 pion and proton quasi-TMDPDFs on the CLS ensemble H101. At present the CS kernel for nonperturbative transverse distances can only be obtained from lattice simulations. Therefore, this is a prime example for combined analyses of experimental and lattice data being needed to obtain relatively complex observables such as TMDPDFs. The fact that we compare for the first time  four qualitatively different cases, namely the TMDPDFs $f_1$ and $e$ of proton and pion, is the primary merit of our investigation. The fact that all four sets of results agree confirms the universality of the CS kernel and suggests that for intermediate transverse distances $0.8\ \mathrm{GeV}^{-1}\lesssim b \lesssim 2.6\ \mathrm{GeV}^{-1}$ systematic errors are not important. Our results are consistent with previous work within uncertainties.  
The main challenge for future work is to quantify all sources of systematic uncertainties. We have demonstrated that combining twist-2 and twist-3 TMDPDFs for different hadrons is conducive to this end.

\begin{acknowledgments}
We acknowledge PRACE for awarding us access to SuperMUC-NG at GCS@LRZ, Germany. The authors thank the Rechenzentrum of Regensburg for providing the Athene Cluster for supplementary computations. We thank the CLS Collaboration for sharing the lattices used to perform this study. The LQCD calculations were performed using the multigrid algorithm~\cite{Babich:2010qb,Osborn:2010mb} and Chroma software suite~\cite{Edwards:2004sx}. We also thank Gunnar Bali, Sara Collins and Christian Zimmermann for helpful discussions. H.T.S., A.S., M.S., L.W. and Y.Y. are supported by a NSFC-DFG joint grant under grant No. 12061131006 and SCHA~458/22. Y.Y. is also supported in part by the Strategic Priority Research Program of Chinese Academy of Sciences, Grant No. XDB34030300 and XDPB15, and also by the National Natural Science Foundation of China (NSFC) under Grants No. 12293062.  A.V. is funded by the \textit{Atracci\'on de Talento Investigador} program of the Comunidad de Madrid (Spain) No. 2020-T1/TIC-20204. A.V. is also supported by the Spanish Ministry grant No. PID2019-106080GB-C21. M.E. is supported by the U.S.~Department of Energy, Office of Science, Office of Nuclear Physics through grant No. DE-FG02-96ER40965 and through the TMD Topical Collaboration.
\end{acknowledgments}

\section*{Appendices}
\appendix

\section{Comparison of the CS kernel extracted from different momentum pairs}
\label{app:momentum-eff}

\begin{figure*}[tbh]
\centerline{
\includegraphics[width=1.0\textwidth]{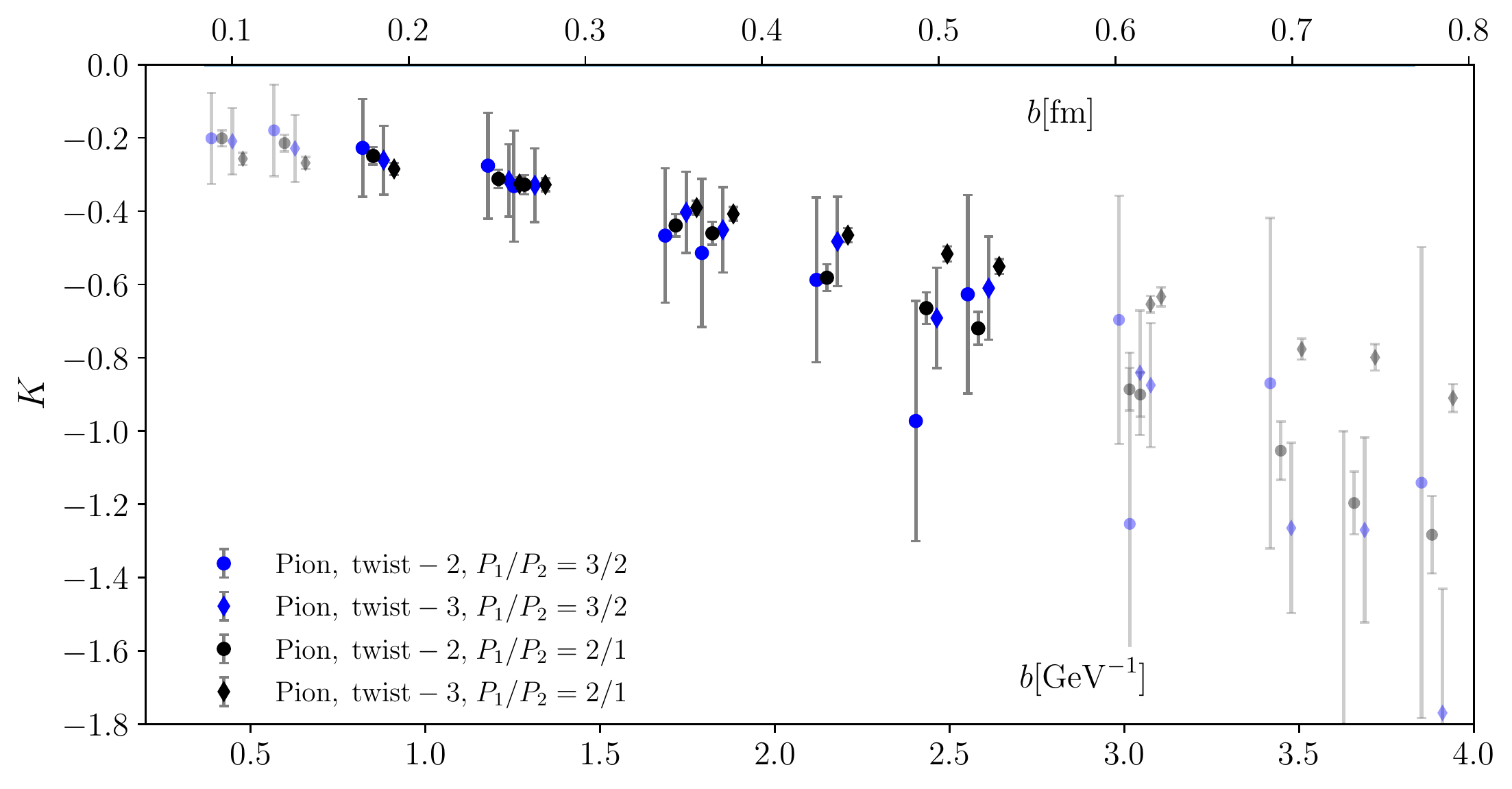}
}
\caption{Comparison of CS kernel extracted from different twists and momentum pairs. The points are slightly shifted horizontally for better visibility.} 
\label{fig-cskernel_all}
\end{figure*}

We compare the extracted CS kernel from different momentum pairs in Fig.~\ref{fig-cskernel_all}. Consistent (within errors) values for the CS kernel can be observed for all $b$ for which the data is reliable. This indicates that the power corrections may be strongly suppressed even at the smallest momentum considered in this work. In addition, the insensitivity of the CS kernel to $M^2/(xP^+)^2$ also supports the rationale of using a large valence quark mass in our calculation {\em a posteriori}.

\begin{figure*}[tbh]
\centerline{
\includegraphics[width=.5\textwidth]{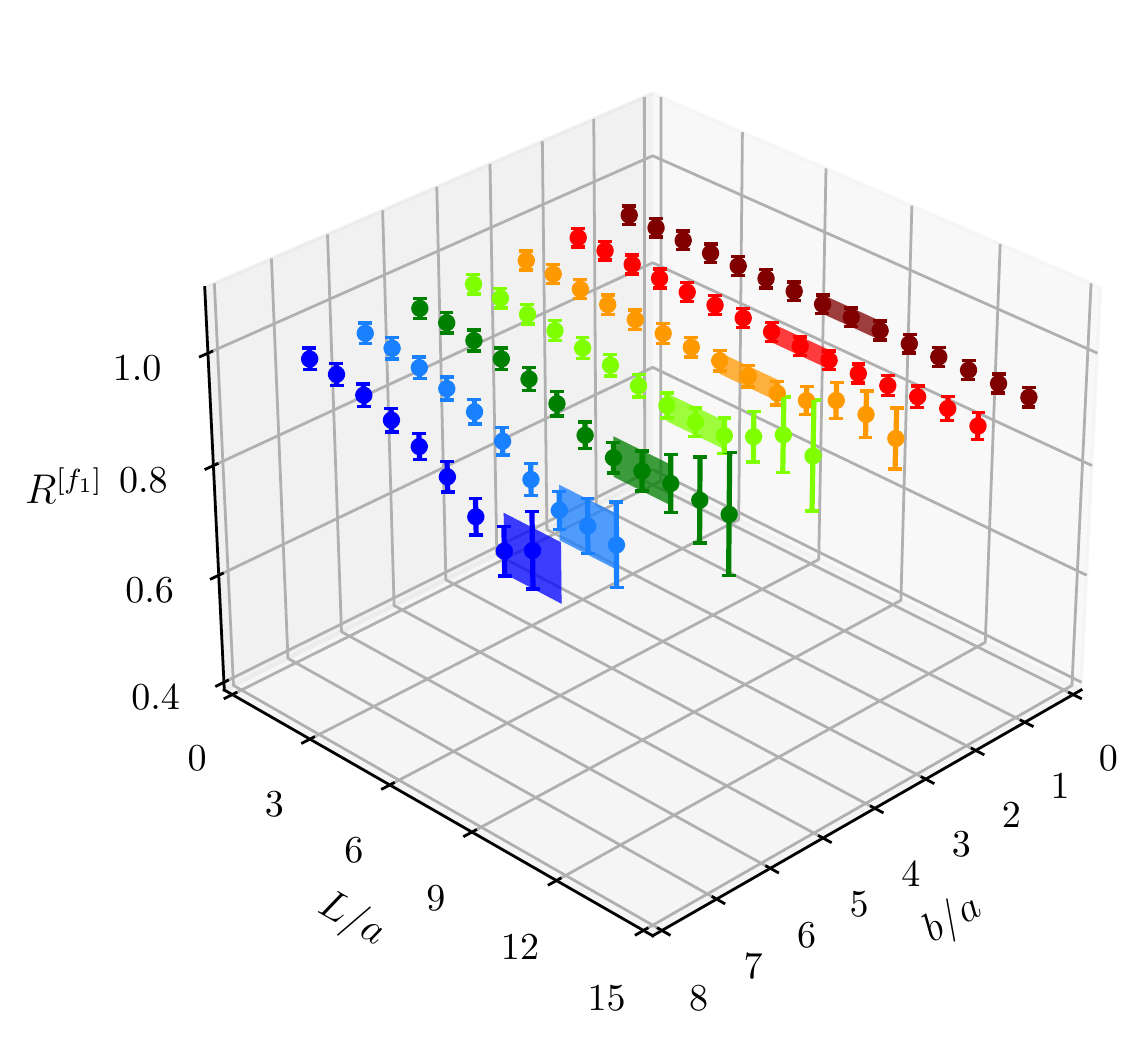}
\includegraphics[width=.5\textwidth]{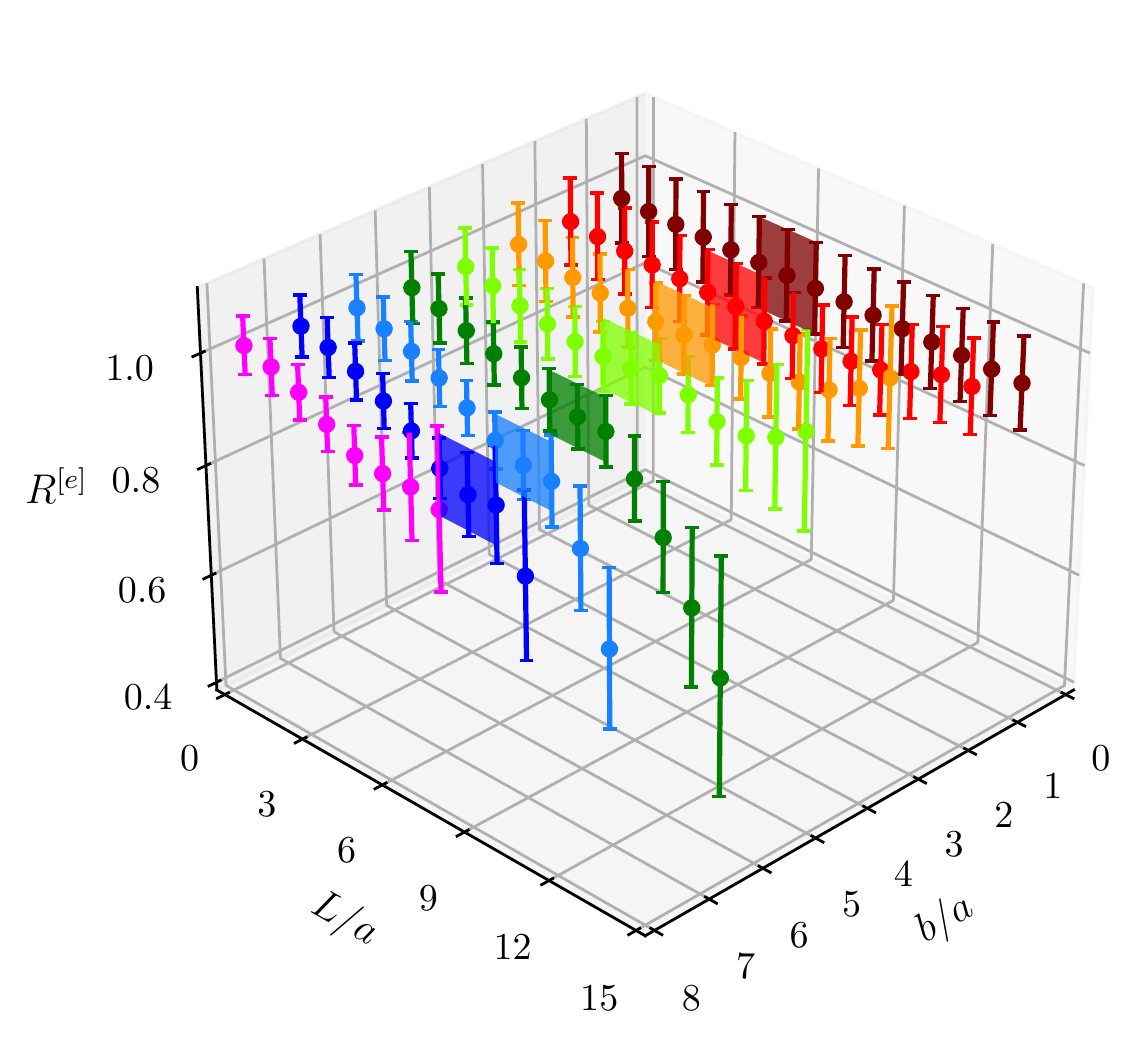}
}
\centerline{
\includegraphics[width=.5\textwidth]{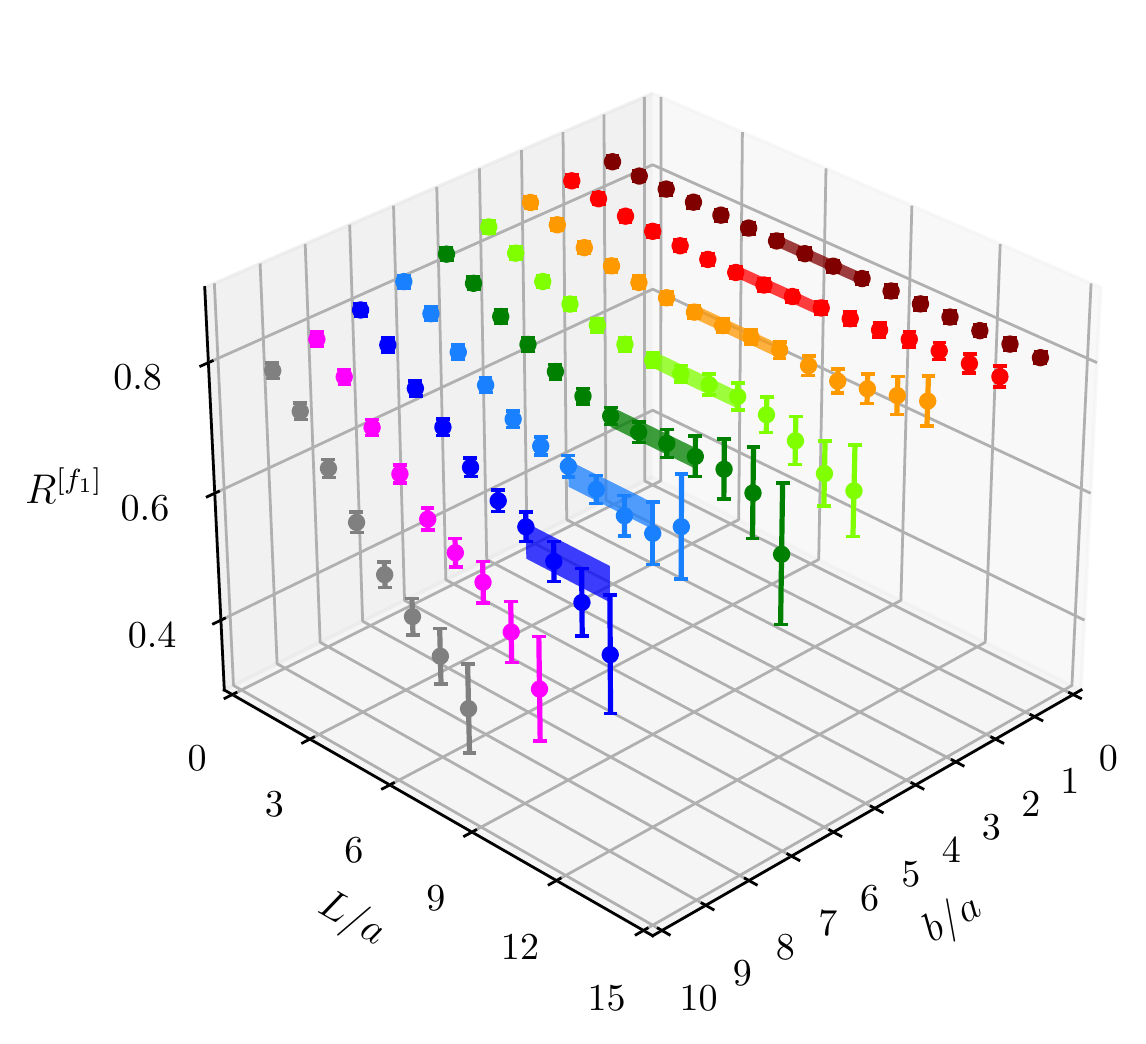}
\includegraphics[width=.5\textwidth]{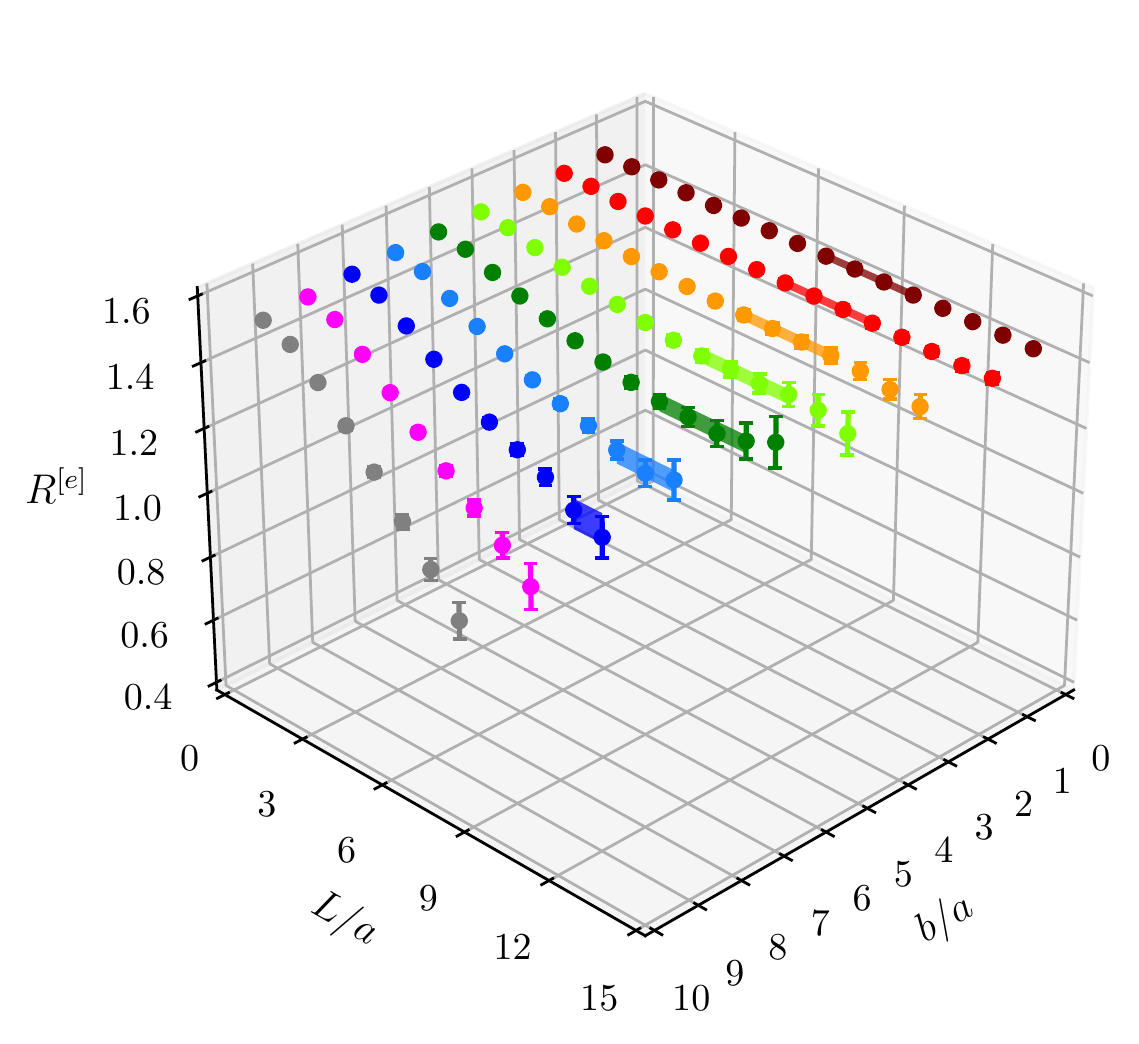}
}
\caption{\textit{Top}: lattice results for the ratios $R^{[f_1]}$ and $R^{[e]}$ in the proton case from the momentum pair $P_1/P_2=3/2$ at different transverse separations $b$. The colored bands indicate the results of constant fits in the $L$-interval $[8a, 10a]$ for $f_1$ and $[6a, 8a]$ for $e$. \textit{Bottom}: Same as above but in the pion case from the momentum pair $P_1/P_2=2/1$ at different transverse separations $b$. The colored bands indicate the results of constant fits in the $L$-interval $[6a, 9a]$ for $f_1$ and $[8a, 11a]$ for $e$. 
}
\label{fig:ratio_more}
\end{figure*} 

\section{More instances of constant fits in $L$ of $R$}
\label{app:more-R}

We first show constant fits of the ratio $R$ in $L$ for the proton in the top panels of Fig.~\ref{fig:ratio_more}. It can be seen that the pattern of change is similar as in the pion case but the fit interval needs to be adjusted. Next, we present the results from the momentum pair $P_1/P_2=2/1$ in the pion case in the bottom panels of  Fig.~\ref{fig:ratio_more}. Compared to the larger momentum case, the plateau appears at larger $L$ for small momentum. We fit the ratio to a constant in the intervals $[6a, 9a]$ for $f_1$ and $[8a, 11a]$ for $e$ at all values of $b$. The plateaus can be identified with much smaller statistical uncertainties, as expected.

\section{A cross check of the determination of $\mathbf{M}$}
\label{app:check-M}
\begin{figure}[tbh]
\centerline{
\includegraphics[width=0.5\textwidth]{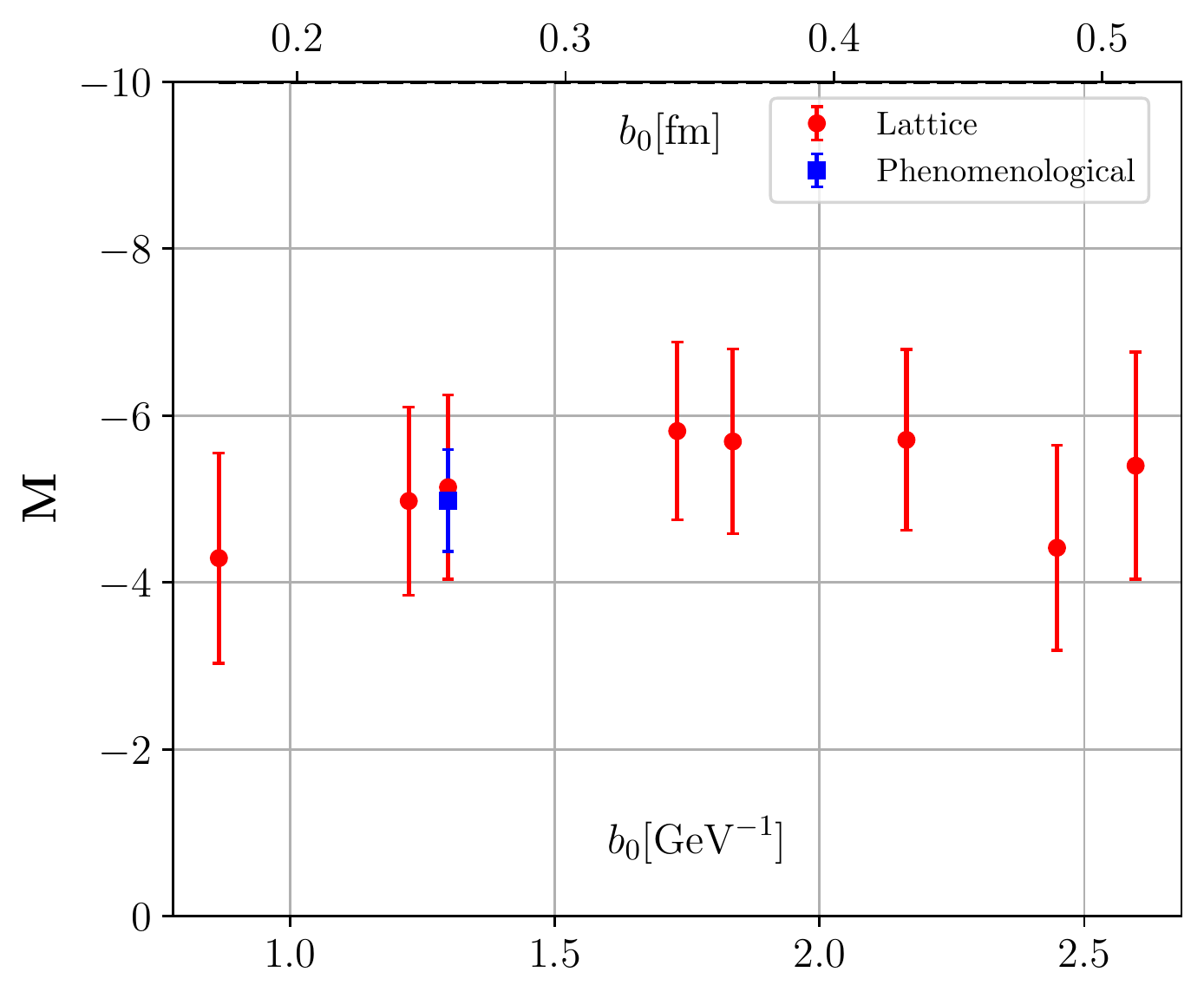}
}
\caption{Comparison of $\mathbf{M}$ determined using phenomenological results and lattice results of the CS kernel.} 
\label{fig-check-M}
\end{figure} 
 $\mathbf{M}$ determined in the main text is based on the (phenomenological) SV19 results. It is shown in Fig.~\ref{fig-check-M} labelled as ``Phenomenological".  $\mathbf{M}$ can also be determined using lattice results of the CS kernel. We take the state-of-the-art lattice results from Ref. \cite{LatticePartonLPC:2023pdv}. As for $R$ we use the twist-3 pion case for a demonstration. $\mathbf{M}$ is determined in the same way as described in the main text with $b_0$ in the range $0.8\ \mathrm{GeV}^{-1}\lesssim b_0 \lesssim 2.6\ \mathrm{GeV}^{-1}$. The error of the lattice determined CS kernel enters via Gaussian bootstrap. The resultant $\mathbf{M}$ is shown in Fig.~\ref{fig-check-M} with label ``Lattice". From the figure we can see that, though with large errors, $\mathbf{M}$ is not sensitive to $b_0$.

\section{Combination of previous lattice results}
\label{app:error-combine}
\begin{figure*}[tbh]
\centerline{
\includegraphics[width=0.5\textwidth]{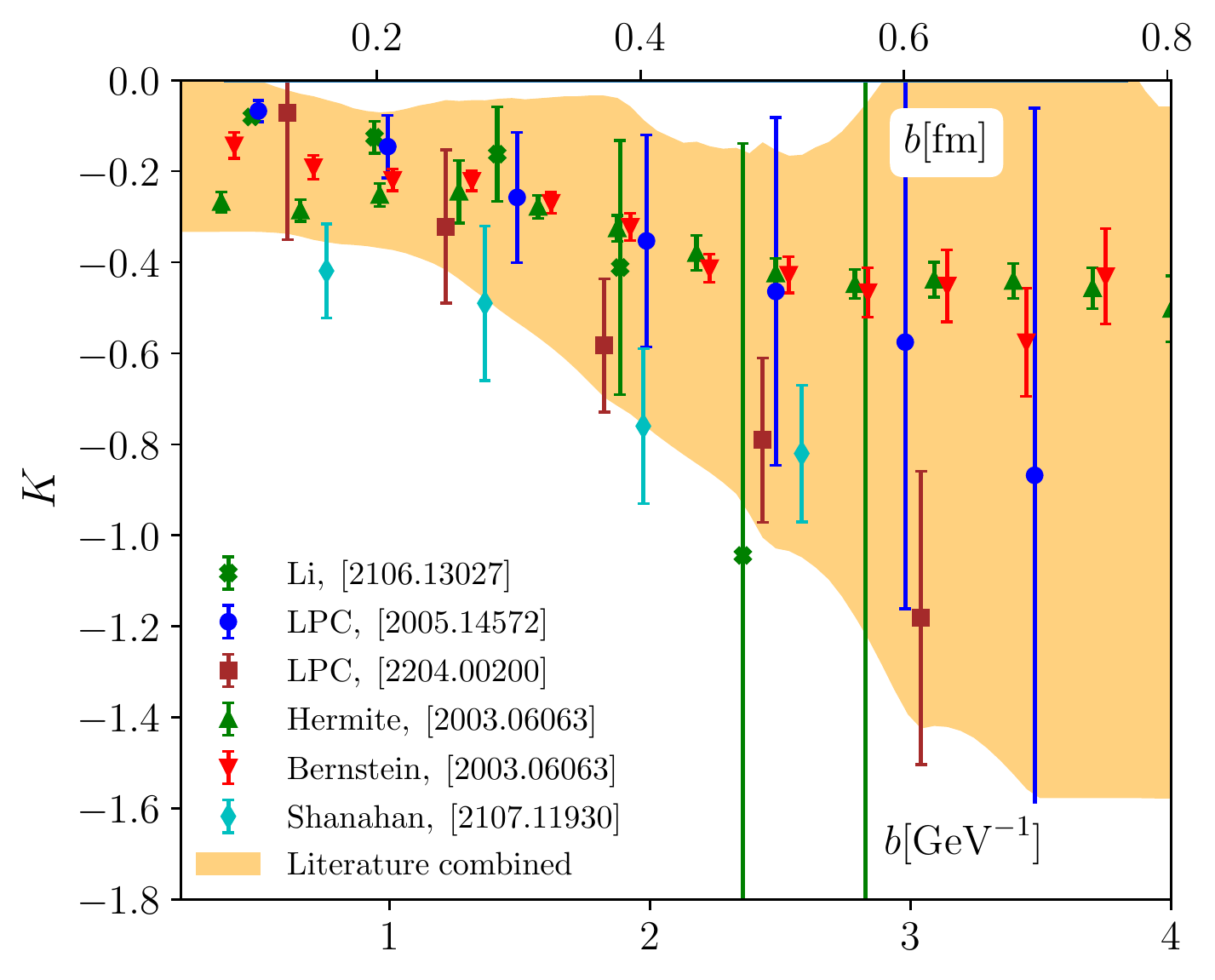}
\includegraphics[width=0.5\textwidth]{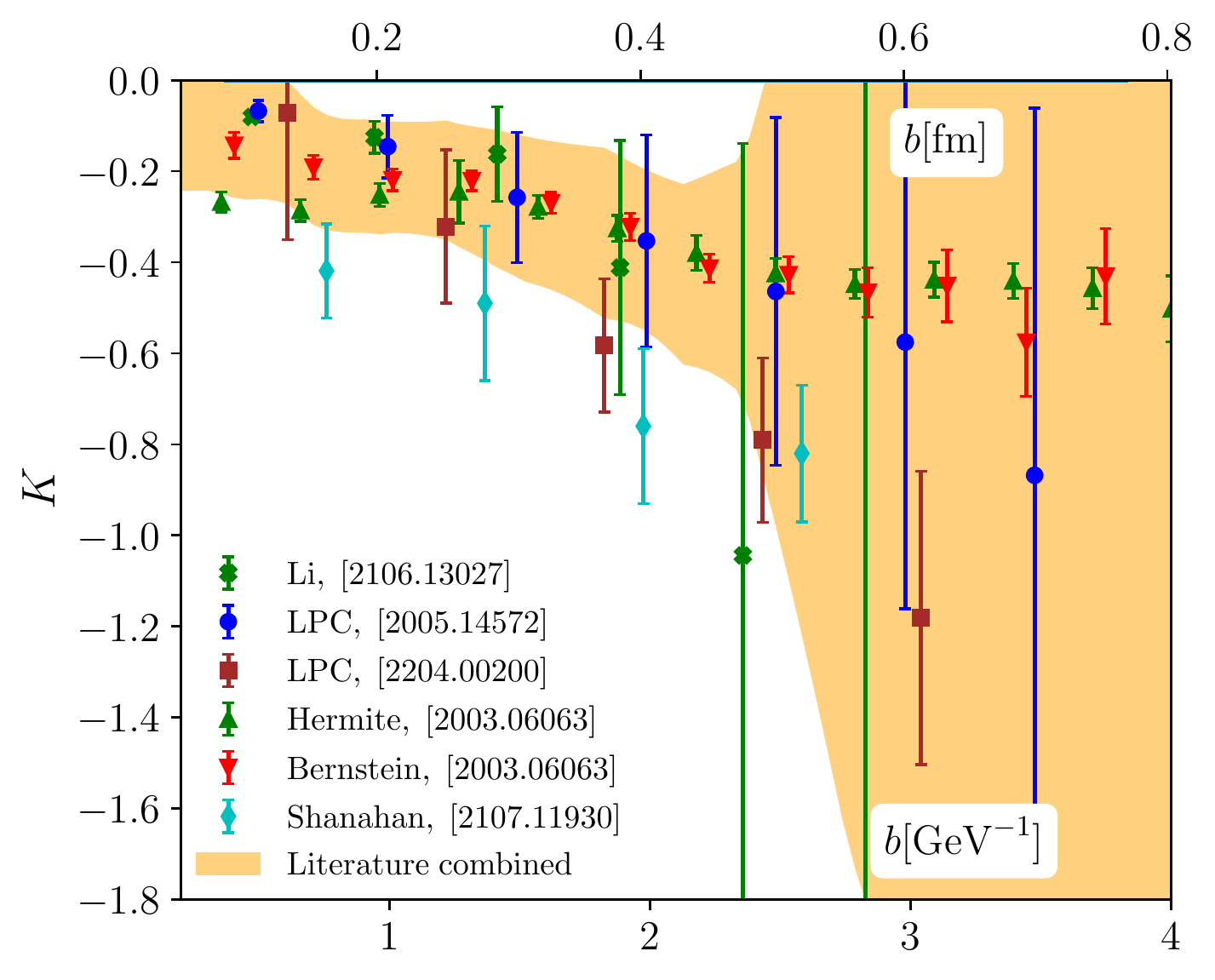}
}
\caption{Combination of previous lattice extractions of the CS kernel \cite{Li:2021wvl, LatticeParton:2020uhz, LPC:2022ibr, Shanahan:2020zxr, Shanahan:2021tst} using Method 1 (left) and Method 2 (right). See main text for details of the two methods.} 
\label{fig:combine-literature}
\end{figure*} 

In the left panel of Fig.~\ref{fig:cskernel_final} we compared our results with the results of previous lattice calculations. Because having all relevant data points in a single plot would not result in a clear presentation of the available information, we summarize all previous lattice results by just one yellow error band. We point out that there are two commonly used methods in the literature to combine different datasets. In the following we dwell on how they are implemented. Note that we treat the error from individual extractions as statistical error, although in some cases it is the sum of the statistical and systematic ones. We have also linearly interpolated between different $b$ values. The first method is the same as the one used in \cite{Altenkort:2023oms}, which we call Method 1: we first generate Gaussian bootstrap samples for each extraction based on its mean and error. Then the samples at the same $b$ from all extractions are combined. The expectation value can be estimated as the median drawn from this pool and the error can be estimated by the 68\% confidence interval. In the left panel of Fig.~\ref{fig:combine-literature} we explicitly present all the previous lattice results and the combined band from Method 1, which is also the one shown in Fig.~\ref{fig:cskernel_final}.

In Method 2 the mean and the statistical and systematic uncertainties are calculated according to
\begin{equation}
\begin{split}
&\langle X \rangle \equiv \Big ( \sum_i \frac{1}{(\delta X_i)^2}\Big )^{-1} \sum_i \frac{X_i}{(\delta X_i)^2},\\
&\delta \langle X \rangle_{stat} \equiv \sqrt{\Big ( \sum_i \frac{1}{\delta X_i^2}\Big )^{-1}},\\
&\delta \langle X \rangle_{syst} \equiv \sqrt{ \big \langle (X-\langle X\rangle)\big \rangle^2},
\end{split}
\end{equation}
where $\delta X_i$ denotes statistical errors of individual extractions and $X_i$ the mean value of the individual extraction. The final error is the sum of the statistical and systematic uncertainty. The results are given in the right panel of Fig.~\ref{fig:combine-literature}.

We remark that in both methods systematic and statistical uncertainties are included. Method 1 is the usual average while Method 2 provides weighted means. Note that in some cases, if the errors of an individual data set are extraordinary small, in Method 2 it will dominate the estimation, for instance the Hermite and Bernstein results. However, previous lattice results usually suffer from various uncontrolled systematics, e.g. the Hermite and Bernstein results are obtained in the quenched approximation and thus sizable systematics from dynamical quarks can be foreseen. In such cases, Method 2 is too aggressive in estimating the uncertainty. For this reason we adopt Method 1 in this work, which gives a broader error band, in the region $0.8\ \mathrm{GeV}^{-1}\lesssim b \lesssim 2.6\ \mathrm{GeV}^{-1}$ to be on the safe side.

\section{A possible estimate for the lattice artifacts}
\label{app:latt-artifact}

Even though it is hard to quantify the lattice spacing effect without adding more simulations, we propose to estimate it in the following, indirect way. It is known that there is a range for $b$ where perturbation theory and the factorization formula are applicable: $\frac{1}{P^+}\ll b \ll \frac{1}{\Lambda_{\mathrm{QCD}}}$. For our lattice calculation this is $0.67\ \mathrm{GeV}^{-1}\ll b \ll 2.7\ \mathrm{GeV}^{-1}$ for the pion and  $0.57\ \mathrm{GeV}^{-1}\ll b \ll 2.7\ \mathrm{GeV}^{-1}$ for the proton. In the manuscript we choose $b=3a=1.3\ \mathrm{GeV}^{-1}$, at which point we normalize our lattice extraction to the phenomenological value from SV19. In principle one can also choose $b=2\sqrt{2}a=1.2\ \mathrm{GeV}^{-1}$ (which is available in this study) where we think we can still trust the phenomenological extraction. There will be an overall shift for the CS kernel determined from these two options. Imagine that now a simulation at a finer lattice spacing is added. We hope that the CS kernel results for the two alternative normalization points differ less for the finer lattice than for the coarser one. Thus the latter can be used as a rough estimate for the lattice spacing effect. Naturally, we also hope that the continuum extrapolated result differs less from the SV19 curve than the present one, based on just one lattice spacing. The resultant systematic uncertainties are shown as outer error bars in the left panel of Fig.~\ref{fig:cskernel_final}.

\bibliographystyle{apsrev4-1}
\bibliography{ref}

\end{document}